\documentclass[10pt]{sigplanconf}

\newtheorem{example}{Example}
\newtheorem{definition}{Definition}
\usepackage{multirow}
\usepackage{fixltx2e}
\usepackage{hyperref}
\usepackage{pstcol}
\usepackage{graphicx}
\usepackage{pstricks}
\usepackage{pst-node}
\usepackage{pst-rel-points}
\usepackage{algorithm}
\usepackage{algorithmic}

\newcommand{\cmt}[1]{} %commented code
\newcommand{\mcspec}[1]{\psframebox[framesep=0,fillcolor=lightgray,fillstyle=solid,linestyle=none]{#1}}
\begin{document}

%
% --- Author Metadata here ---
\conferenceinfo{APPLC'13}{February 23, 2013, ShenZhen, China}
\CopyrightYear{2013} % Allows default copyright year (2002) to be over-ridden - IF NEED BE.
%\crdata{}
% --- End of Author Metadata ---

\title{Retargeting GCC: Do We Reinvent the Wheel Every Time?}

\authorinfo{Saravana Perumal P}
           {Department of CSE, IIT Kanpur}
           {\url{saravanan1986@gmail.com}}
\authorinfo{Amey Karkare}
           {Department of CSE, IIT Kanpur}
           {\url{karkare@cse.iitk.ac.in}}

\maketitle
\begin{abstract}
%%The availability of GCC ports for a large number of targets
%%stands testimony to the success of the retargeting model of
%%GCC.  
Porting GCC to new architecture requires writing a
Machine Description (MD) file that contains mapping from
GCC's intermediate form to the target assembly code.
Constructing an MD file is a difficult task
%%, partly because
%%of its large size, but mainly 
because it requires the user to
understand both (a) the internals of GCC, and (b) the
intricacies of the target architecture. 
%%While different
%%architectures have different instruction sets, these
Instruction sets of different architectures exhibit
significant amount of semantic similarities across a large
class (for example, the instruction sets for RISC
architectures) and differ only in syntax.  Therefore, it is
expected that MD files of machines with similar architectures
should also have similarities. To confirm our hypothesis, we
created {\em mdcompare}, a tool to (a) extract RTL patterns
(machine independent abstraction of RTL templates) from MD
files of well known architectures and (b) compare the
similarity of patterns across architectures.  The results are
encouraging; we found that 28\% -- 70\% RTL expressions are
similar across pairs of MD files, the similarity percentage
being on the higher side for pairs of similar architectures.

%% The tool forms the first  step towards our long term goal: to
%% build a tool that can automatically generate major part of an
%% MD file  for a  new architecture from  existing MD  files for
%% similar architectures.   We believe  that this will  make the
%% process of retargeting GCC simpler and less error prone as it
%% will reduce the user involvement and allow reuse of existing,
%% well tested patterns.
\end{abstract}


\section{Introduction}
\label{sec:intro}

The GNU Compiler Collection (GCC)~\cite{gcc-web} is an
integrated distribution of compilers for several programming
languages~\cite{usinggcc}.  GCC
%\footnote{In the rest of the
%  paper, we use GCC as a singular noun. However, it should be
%  understood that we are talking about a collection of
%  compilers that are part of the same framework} 
is the most widely used compiler collection for developing
applications that run across several different architectures
and operating systems. One of the strengths of GCC is that it
is highly portable, owing to the fact that the core compiler
part of GCC does not have any machine-specific code, but has
parameters which depend on the target machine's
features~\cite{gccint}. The information about the target
machine is obtained from {\em Machine Description (MD)}
files. An MD file is a text file containing mappings from
GCC's intermediate representation to the instruction set of a
target architecture. This is an elegant way of alienating the
machine-specific details from the core compiler.

%\subsection{Motivation}
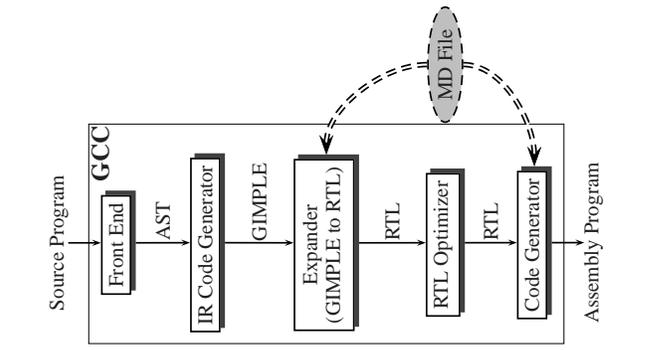
\begin{figure}[t]
\begin{center}
\newcommand{\gccbox}{\mbox{
\scalebox{0.79}{
\psset{unit=1mm}
\begin{pspicture}(0,0)(40,95)
%  \psframe(0,0)(55,95)

  \psframe[linewidth=.1pt](-2,8)(35,88)
  \putnode{source}{origin}{15}{93}{Source Program}
  \putnode{fe}{source}{0}{-10}{\psshadowbox{Front End}}
  \ncline{->}{source}{fe}

  \putnode{gcc}{fe}{15}{3}{\large\bf GCC}
  \putnode{ircg}{fe}{0}{-15}{\psshadowbox{IR Code Generator}}
  \ncline{->}{fe}{ircg}  \Aput[0.1]{AST}

  \putnode{exp}{ircg}{0}{-20}{\psshadowbox{%
      \begin{tabular}{@{}c@{}}
        Expander \\
        (GIMPLE to RTL)
      \end{tabular}}}
  \ncline{->}{ircg}{exp}   \Aput[0.1]{GIMPLE}

  \putnode{ropt}{exp}{0}{-20}{\psshadowbox{RTL Optimizer}}
  \ncline{->}{exp}{ropt} \Aput[0.1]{RTL}

  \putnode{cg}{ropt}{0}{-15}{\psshadowbox{Code Generator}}
  \ncline{->}{ropt}{cg} \Aput[0.1]{RTL}

  \putnode{asm}{cg}{0}{-10}{Assembly Program}
  \ncline{->}{cg}{asm}

  \putnode{mdf}{exp}{30}{-20}{\psovalbox[%
      fillcolor=lightgray,fillstyle=solid,linestyle=dashed]{MD File}} 
  \nccurve[angleA=90,doubleline=true, linestyle=dashed]{->}{mdf}{exp}
  \nccurve[angleA=-90,doubleline=true, linestyle=dashed]{->}{mdf}{cg}

\end{pspicture}}}}

\rotatebox{90}{%
        \gccbox
}
\caption{High Level Flow of GCC\label{fig:gccflow}}
\end{center}
\vskip -4mm
\end{figure}

Figure~\ref{fig:gccflow} shows a high level view of
translation of source code to target machine code by GCC.
The front end parses the source language, converts it to an
intermediate representation called GIMPLE. After a few
internal transformations, GIMPLE representation is converted
to another intermediate representation called RTL (Register
Transfer Language). The RTL representation is converted to
the target assembly code. The conversion from GIMPLE to RTL
and from RTL to assembly is guided by the templates present
in MD file. These are called {\em RTL expressions} and encode
all machine-specific information.  The MD file is used by the
GCC framework during the building of the GCC compiler.

Constructing the MD file for a target architecture forms the
most important step in porting GCC. Writing an MD file for a
new architecture needs a good understanding of GCC's
intermediate representations, the RTL expressions, and the
instruction set of the target architecture.  Given the
complexity of the modern architectures and the variety of
architectures available, this is a huge ask. 
So, in practice, an MD file for a new architecture is constructed
from an MD file of a similar architecture by making
modifications to suit the needs. This is a method of
trial-and-error, and construction of MD file in this way is
observed to be complex, verbose and repetitive~\cite{Khedker,
  kai-wei, Sameera}. Any mistake in MD file may result in the
compiler producing wrong or worse, inefficient code, without
the user detecting it quickly.  This is because the MD
files themselves are huge (running into few thousand
lines for typical architectures). For example, MD file for ARM
has 30,943 lines, for i386 has 38,817 lines, and MIPS has
15,534 lines for GCC version 4.6.1~\cite{gcc-web}.

Retargeting GCC could be simplified significantly if the
process of writing MD files is fully automated. However, due
to the complexity of the problem, this is virtually
impossible.  A middle way to solve this problem is to
partially automate the way user do it today. In other words,
to create an MD file for a new architecture, use the MD file
for a similar architecture and automate the generation of
parts that correspond to the similar features in the
architectures. Rest of the parts can be provided by the
user. The hypothesis here is that {\em similar
  instructions for two architectures have similar RTL
  expressions in the respective MD files}.

In this paper, we describe our experiments to verify this
hypothesis and report our findings. We first present an example to
motivate the problem and to explain the notion of similarity
used by us.

\subsection{A Motivating Example}
Consider the problem of adding two numbers on a given
architecture. The assembly instruction to be used depends on
the type of numbers to add (SI integer, DI integers, floating
points, sign extension required etc.), the type of storage
(register, memory, immediate, etc.) etc. We give an example
of the similarity that exists across MIPS and ARM
architectures for addition instruction\footnote{The RTL
  expressions in this paper are modified for ease of
  explanation and to avoid referring to complex concepts that
  are out of scope of this work.}.

\begin{figure}[t!]
\begin{center}
\scalebox{.75}{
  \renewcommand{\arraystretch}{1}
  \begin{tabular}{|c|@{}c@{}|} \hline
    {\bf Arch.} & {\bf RTL Expression}
    \\ \hline \hline
    MIPS &
    \textrm{%
      \begin{tabular}{@{}l@{}}
        (define\_expand ``add$\langle$mode$\rangle$3'' \\
        \ [(set \mcspec{(match\_operand:GPR 0 ``register\_operand'')} \\
          \ \rule{8mm}{0pt} (plus:\mcspec{GPR} \mcspec{(match\_operand:GPR 1 ``register\_operand'')} \\
          \ \rule{27mm}{0pt}\mcspec{(match\_operand:GPR 2
          ``arith\_operand'')}))] \\
        \ \ldots)
      \end{tabular}
    }
    \\ \hline
    ARM &
    \textrm{%
      \begin{tabular}{@{}l@{}}
        (define\_expand ``addsi3'' \\
        \ [(set \mcspec{(match\_operand:SI 0
            ``s\_register\_operand'' ``'')} \\
	  \ \rule{8mm}{0pt} (plus:\mcspec{SI}
          \mcspec{(match\_operand:SI 1
            ``s\_register\_operand'' ``'')} \\
	  \ \rule{27mm}{0pt} \mcspec{(match\_operand:SI 2
            ``reg\_or\_int\_operand'' ``'')}))]\\
        \ \ldots)
      \end{tabular}
    }
    \\ \hline
    \multicolumn{2}{p{1.25\columnwidth}}{(a) Similar RTL
      Expression for addition instruction for MIPS and ARM 
      architectures. Machine-specific parts are shaded.} 
  \end{tabular}
}
\end{center}

\begin{center}
\scalebox{.9}{
\renewcommand{\arraystretch}{1.2}
\begin{tabular}[b]{|@{}c@{}|@{}c@{}|} \hline
  {\bf RTL Pattern} & {\bf Tree Representation } \\ \hline 
  \textrm{
    \begin{tabular}[b]{@{}l@{}}
      %      (define\_expand ``addsi3'' \\
      [(set\ \mcspec{\$arg0} \\
        \ \ \ \ \ (plus:\mcspec{\$mode} \mcspec{\$arg1} \\
        \ \ \ \ \ \ \ \ \ \ \ \ \ \ \ \ \ \ \ \mcspec{\$arg2}))] \\
      \\
    \end{tabular}
  }
  & \scalebox{.95}{
    \psset{unit=1mm}
    \begin{pspicture}(0,0)(42,25)
      %\psframe(0,0)(42,25)
      \putnode{n0}{origin}{16}{21}{\pscirclebox{set}}
      \putnode{n1}{n0}{-10}{-9}{\psovalbox[framesep=0]{\mcspec{\$arg0}}}
      \putnode{n2}{n0}{10}{-9}{\psovalbox[framesep=0]{plus:\mcspec{\$mode}}}
      \putnode{n3}{n2}{-10}{-9}{\psovalbox[framesep=0]{\mcspec{\$arg1}}}
      \putnode{n4}{n2}{10}{-9}{\psovalbox[framesep=0]{\mcspec{\$arg2}}}
      \ncline[nodesepA=-1,nodesepB=-.5]{-}{n0}{n1}
      \ncline[nodesepA=-1]{-}{n0}{n2}
      \ncline[nodesepB=-.5]{-}{n2}{n3}
      \ncline{-}{n2}{n4}
  \end{pspicture}} \\ \hline
    \multicolumn{2}{c}{\scalebox{.85}{(b) Common RTL pattern for addition.}}
  \end{tabular}
}
\end{center}

\begin{center}
\scalebox{.9}{\small
\renewcommand{\arraystretch}{1.2}
\begin{tabular}[b]{|@{}c@{}|p{30mm}|p{33mm}|} \hline
  {\bf Parameter} & \multicolumn{2}{c|}{\bf Value} \\ \cline{2-3}
  & MIPS & ARM  \\ \hline \hline
  \$mode & GPR & SI \\ \hline
  \$arg0 & (match\_operand:GPR 0 ``register\_operand'') & (match\_operand:SI 0 ``s\_register\_operand'' ``'') \\ \hline
  \$arg1 & (match\_operand:GPR 1 ``register\_operand'') & (match\_operand:SI 1 ``s\_register\_operand'' ``'') \\ \hline
  \$arg2 & (match\_operand:GPR 2 ``arith\_operand'') & (match\_operand:SI 2 ``reg\_or\_int\_operand'' ``'')  \\  \hline
  \multicolumn{3}{c}{\scalebox{.85}{(c) Instantiation of common RTL pattern.}}
  \end{tabular}
}
\end{center}

\vskip -2mm
\caption{Use of RTL pattern across multiple architectures\label{fig:motiv}}
\vskip -3mm
\end{figure}

\begin{example}\label{exmp:motiv}
Figure~\ref{fig:motiv}(a) shows RTL expressions to add two
numbers on MIPS and ARM architectures respectively. Without
going into the details of the semantics, we notice that the
form of the RTL expression ({\tt [(set \ldots (plus
    \ldots))]}) to select the appropriate instruction is
identical for these architectures. We call this common form,
that is obtained by abstracting out the machine-specific
parts of an RTL expression, an {\em RTL
  Pattern}.

 Figure~\ref{fig:motiv}(b) shows the corresponding
RTL template, the parameters to the templates, and the
instantiation of parameters to recover the original
expression. It also shows the tree representation of the RTL
pattern to help visualize it.
\end{example}
 
The pattern in this case is extracted automatically by our
tool by looking at RTL expressions present in the MD
files. It is a part of the set of arithmetic patterns
required for similar architectures. While retargeting GCC for
a similar architecture, this is one of the arithmetic
patterns that needs to be filled with machine-specific
values.

We generalize this observation as follows: if we can find a
set of minimal RTL patterns, that are common across machines
having similar architecture, the process of retargeting can
be made simpler and systematic where parts of MD files for
new architectures can be generated automatically. Our
hypothesis simply says that the RTL expressions are alike for
two instructions that behave alike.
 
To verify our hypothesis, we experimented with five well
known architectures (ARM, i386, MIPS, SPARC and VAX) for whom
the MD files are present in the GCC's source tree. We
developed a tool {\em mdcompare} to compare similarities
across these MD files. To do so, we identified patterns that
are common across machines and measured the percentage of RTL
expressions that can be generated by these patterns by
supplying the machine-specific information. The results
show up to 70\% similarity for similar architectures. 
The results largely confirm our hypothesis and justify the
need for an automatic tool to generate MD files for new
architectures from existing MD files for similar architectures.  
\vskip -5mm
\subsection{Contribution of This Paper}
The main contributions of our work are as follows:
\begin{itemize}
\item We describe what are RTL patterns and how these
  are useful
  (Section~\ref{sec:pattern}).
\item We describe the tool {\em mdcompare} that is used to
  extract RTL patterns (machine independent abstraction of
  RTL expressions) from MD files (Section~\ref{sec:anal-md}).
\item The tool is used to compare similarity across pairs of
  5 well known architectures (Section~\ref{sec:expr-pat},~\ref{sec:simmd}).
\end{itemize}

%% \subsection{Organization of the Paper}
%% The rest of this paper is organized as follows. We define RTL
%% pattern and describe its usefulness in
%% Section~\ref{sec:pattern}.  Section~\ref{sec:results}
%% describes the experiments performed on MD files, presents the
%% results and analyzes the important observations. We describe
%% the related work in Section~\ref{sec:related} and conclude
%% the paper in Section~\ref{sec:concl}, giving the directions
%% for future of our project.

\section{RTL patterns and their usage}
\label{sec:pattern}
%% We now describe the RTL patterns, how we identify them and
%% why they are useful. 

%% \subsection{Extracting the Form of an RTL Expression}

GCC compiler follows the model proposed by Davidson and
Fraser~\cite{Davidson} for code generation. Even though the
RTL expressions in an MD file represent machine instructions
for a specific machine, their form is machine-independent. We
call this machine independent form {\em RTL pattern} and use
it for comparing similarity across MD files for different
architectures.
\vskip -2mm
\begin{definition}
An {\em RTL pattern} is an RTL expression whose
machine-specific details are replaced by named parameters.
\end{definition}
\vskip -2mm
An RTL pattern can correspond to more than one RTL
expressions that differ only in machine-specific
details. These RTL expressions could be for same machine or
for different machine. Figure~\ref{fig:motiv}(b) shows an
RTL pattern that can give rise to RTL expressions in
Figure~\ref{fig:motiv}(a) by providing suitable values to
parameters \$mode, \$arg0, \$arg1 and \$arg2.

\vskip -2mm
\paragraph{RTL Patterns}
Machine-specific parts occur at well defined places in
various RTL expressions, hence it is easy to extract RTL
patterns from expressions. For example, the mode of an
arithmetic operator (e.g., plus) is not part of a pattern
because the modes supported are machine specific. Similarly,
the match\_operand expressions are machine-specific as they
have machine-specific fields like predicate and constraint.
Such details are removed from RTL expression and replaced by
named parameters to obtain an RTL pattern. It should be
obvious that an RTL pattern, when instantiated with suitable
machine-specific parameters like mode or match\_operand
expressions, will result in an RTL expression.

Our tool {\em mdcompare} parses MD files and stores the RTL
expressions as expression trees. It generates RTL patterns
also in tree form. To extract the pattern from an expression,
the tool traverses the expression tree in a bottom up fashion
and replaces machine-specific subexpression trees by
parameterized subpattern trees.  It also maintains a mapping
between the parameter variables and the corresponding values
to allow reuse of parameters. This is important to keep the number of
parameters small as some machine-specific features like mode
(SI, DI, GPR etc.) are used several times in a single RTL
expression.  The tool maintains a list of patterns, already
identified, sorted in the increasing order of their
heights. When a new (sub)pattern is encountered, it is
compared with patterns of same height for equality and, if
not already present, is added to the list.  We use the height
of the patterns to avoid unnecessary comparison between
patterns of different heights.

\vskip -2mm
\paragraph{Usefulness of  RTL Patterns}
We found that RTL patterns are useful in two ways: (a)
RTL patterns help in understanding the
structure of MD files by allowing the user to focus on the
basic forms of instructions, without worrying about such
RTL Patterns Help in Finding Similarities Across
  MD Files. This is explained in details in next section.

\section{Experiments and Results}
\label{sec:results}
We have developed a tool {\em mdcompare} to experiment with
MD files. We now describe our experience with the tool and
explain the results obtained.

\subsection{Analyzing MD Files with mdcompare\label{sec:anal-md}} 
The tool {\em mdcompare} parses MD files and stores the RTL
expressions in tree form. The components of MD files
that are not of interest to create RTL pattern, for
e.g. Output Patterns~\cite{gccint}, are ignored.

Once the expression tree is generated the tool can
systematically remove machine independent parts of the expression
to compute RTL patterns and the number of occurrences. These
RTL patterns can be used to understand the contents of MD
files or to compute similarity of two MD files. We have added
some auxiliary features to the tool to test its correctness
and to enhance its usability. 
\cmt{{Some of the features
implemented are:
\begin{itemize}
\item The tool can split RTL expressions into RTL patterns
  and machine-specific parameters and save them to text
  files. While storing machine-specific parameters, we also
  store the unique id (generated by the tool) of the RTL
  pattern to which it correspond to. Note that more than one
  set of parameters can correspond to same pattern to give
  multiple expressions.This splitting is important because it
  simplifies retargeting by allowing user to manipulate RTL
  templates and parameters independently. This may not be
  very useful now, but it will help in automating the
  retargeting process.
\item The tool can read RTL patterns from one file and
  associated machine-specific parameters from other file in order to   combine them to generate a complete MD
  file. Again, this feature will be useful for automated
  retargeting. We used this feature to test the splitting
  performed by our tool by regenerating the MD file and
  making sure that no expression is missed or generated
  incorrectly.
\item The tool can combine multiple RTL pattern files,
  picking up only those patterns that occur more than some
  user-specified number of times. This feature is useful to
  find commonly occurring patterns 
\end{itemize}
More }} The details of the tool and its feature are given
in details in Saravana's thesis~\cite{saravana2012retarget}.

%\subsection{Results}

To verify our hypothesis that there is a lot of similarity
across MD files, we used the MD files of
five architectures, namely ARM, i386, MIPS, SPARC, VAX 
taken from the back-end of GCC version 4.6.1~\cite{gcc-web}.
Note that ARM, MIPS and SPARC are RISC architectures, while
i386 and VAX are CISC architectures.

\subsection{Expressions and Patterns\label{sec:expr-pat}}
Table~\ref{tab:Pat&Templates} lists the number of RTL
expressions  considered in each of the architecture's machine
description file and the number of patterns that form the
basis of it.

\begin{table}[t]
 \centering
 \caption{Number of RTL Expressions and Patterns.
   \label{tab:Pat&Templates}}
\scalebox{.9}{\small
\renewcommand{\arraystretch}{0.85}
\begin{tabular}{|r | c | c| r |}
\hline
{\bf Arch.} & {\bf\#Expr. (E)} & {\bf\#Patterns (P)} &
{\bf Average (E/P)} \\
\hline
ARM & 1581 & 362 & 4.37\\
\hline
MIPS & 736 & 209 & 3.52\\
\hline
SPARC & 701 & 187 & 3.74\\
\hline
i386 & 2238 & 547 & 4.09\\
\hline
VAX & 125 & 64 & 1.95\\
\hline
\end{tabular}}
\vskip -5mm
\end{table}

From Table~\ref{tab:Pat&Templates}, it can be seen that on
average a pattern is used in 3--4 expressions for an
architecture.  This shows that patterns can help tells us
about the level of redundancy that is present within machine
description files. VAX seems to be an anomaly. This can be
attributed to the CISC nature of VAX architecture because it 
has a small number of instructions with little or no
variations.

\subsection{Similarity Between Machine Descriptions\label{sec:simmd}}
We now describe the results for MD file
similarities. Table~\ref{tab:Inter1} lists the number of
patterns that are common between the architectures and their
percentage. The percentage is computed as follows: Let
architecture $m_1$ have $p_1$ patterns and $m_2$ have $p_2$
patterns. Let $p$ be the number of identical patterns. Then,
the \% similarity of patterns = $\frac{2\times p}{p_1 +
  p_2}\times 100$.
\begin{table}[t!]
  \centering
  \caption{Number and percent of common patterns between pairs of MD files.
    \label{tab:Inter1}}
\scalebox{.9}{\small
\renewcommand{\arraystretch}{1.1}
  \begin{tabular}{|@{\ }r@{\ }|@{\ }c@{\ }|@{\ }c@{\ }|@{\ }c@{\ }|@{\ }c@{\ }|}
    \hline
        {\bf Arch. } & MIPS & SPARC & i386 & VAX \\
        \hline                        
        ARM      & 75 (26.27\%)   & 79 ({\bf 28.78\%}) & 101 (22.22\%)  &
        35 (16.43\%)\\
        \hline                        
        MIPS    &      & 48 (24.24\%)   & 73 (19.31\%)   & 29 (21.25\%)\\
        \hline                        
        SPARC   &      &       & 63 (17.17\%)   & 30 (23.90\%)\\
        \hline                        
        i386    &      &       &      & 34 ({\bf 11.13\%})\\
        \hline
        \multicolumn{5}{@{}p{1.1\columnwidth}@{}}{\small Number
          in each cell denote identical patterns. Numbers in
          parentheses denote the percentage similarity based
          on patterns.}
  \end{tabular}}
\vskip -5mm
\end{table}

From Table~\ref{tab:Inter1}, we can see that 11--29\% of
patterns are common across architecture. These numbers are
not very encouraging to justify our hypothesis. However, on
further investigation we found that the number of expressions
generated by common patterns form a significant percentage of
the MD files. We measures the percentage of the expressions
generated by common patterns as follows: Let architecture
$m_1$ has $e_1$ expressions and $m_2$ has $e_2$ expressions. Let
$p$ be the number of identical patterns, $e_1'$ be the number
of expressions in $m_1$ generated by the $p$ common patterns, and
$e_2'$ be the same number for $m_2$ . Then, the \%
similarity of MD files =
$\frac{(e_1' + e_2')}{e_1 +  e_2} \times
100$. Table~\ref{tab:Inter2} lists the numbers for similarity
based on RTL expressions derived from common patterns. 

\begin{table}[t!]
  \centering
\caption{Number and percent of expressions generated by
  common patterns for pairs of MD files.\label{tab:Inter2}}
\scalebox{0.82}{\small
\renewcommand{\arraystretch}{1.3}
  \begin{tabular}{|@{}r| @{\ }c@{\ } | @{\ }c@{\ }| @{\ }c@{\ } | @{\ }c@{\ } |}
\hline
 {\bf Arch.} & MIPS           & SPARC          &  i386           & VAX            \\\hline
ARM          & 1486 (64.13\%) & 1475 ({\bf 64.63\%)} &  2281 (59.72\%) &  799 (46.83\%) \\ \hline
MIPS         &                &  838 (58.31\%) &  1584 (53.26\%) &  355 (41.23\%) \\ \hline
SPARC        &                &                &  1441 (49.03\%) &  410 (49.63\%) \\ \hline
i386         &                &                &                 &  719 ({\bf 30.42\%}) \\ \hline

\multicolumn{5}{@{}p{1.15\columnwidth}@{}}{Number
  in each cell denote identical patterns. Numbers in
  parentheses denote the percentage similarity based
  on patterns.}
\end{tabular}}
\vskip -5mm
\end{table}

From Table~\ref{tab:Inter2}, it can be seen that 30--65\% of
the expressions share common patterns. It is interesting to
note that VAX architecture does not show much similarity to
others, and derives only a small percentage of expressions
from common patterns.  We believe the reason is that VAX is
the smallest architecture in terms of the number of RTL
expressions: VAX has just 125 RTL expressions resulting in
64 patterns. Further, being a CISC architecture, the
instruction set differs considerably from others. In
contrast, i386 has 2238 expressions and 547 patterns. Many
i386 instructions are similar to those present in RISC
architectures. This is the reason why i386 has large number
of common expressions with RISC architectures, despite itself
being a CISC architecture. If we look at architectures of
relatively similar size (in terms of count of RTL
expressions), the pair MIPS-SPARC has about 60\% of the RTL
expressions that can be instantiated from their common
patterns.

Another interesting study that we performed was, given MD
files for two architectures, say $s$ (source) and $t$
(target), how many RTL expressions of $t$ can be generated by
providing machine-specific information of $t$ to the RTL
patterns of $s$. This study gives us an idea about the
percentage of MD file for $t$ that can be generated
automatically from $s$, provided we have an automatic tool as
powerful as human developers. 

\begin{table}[t!]
 \centering
 \caption{RTL expressions for Target architectures as generated
   from RTL patterns of Source Architectures.
   \label{tab:Inter3}}
 \scalebox{.9}{
\small
\renewcommand{\arraystretch}{1.1}
   \begin{tabular}{|@{\ }l@{\ }| @{\ }p{12mm}@{\ } |@{\ }p{12mm}@{\ }| @{\ }p{12mm}@{\ } | @{\ }p{12mm}@{\ } | @{\ }p{12mm}@{\ } |}            \hline
     Target $\longrightarrow$          & ARM           & MIPS          & SPARC         & i386           & VAX          \\ 
     (Total \#Exprs)                   & (1581)        & (736)         & (701)         & (2238)         & (125)        \\ \cline{1-1}  
     Source $\downarrow$               &               &               &               &                &              \\ \hline \hline
     ARM                               &               & 504 (68.48\%) & 483 (68.91\%) & 1196 (53.44\%) & 88 ({\bf 70.40\%}) \\ \hline
     MIPS                              & 982 (62.11\%) &               & 391 (55.77\%) & 1075 (48.03\%) & 80 (64.00\%) \\ \hline
     SPARC                             & 992 (62.75\%) & 447 (60.73\%) &               &  994 ({\bf 44.41\%}) & 74 (59.20\%) \\ \hline
     i386                              &1085 (68.63\%) & 509 ({\bf 69.16\%}) & 447 (63.76\%) &                & 87 (69.60\%) \\ \hline
     VAX                               & 711 (44.97\%) & 275
     (37.36\%) & 336 (47.93\%) &  632 ({\bf 28.23\%}) &
     \\ \hline
     \multicolumn{6}{@{}p{1.1\columnwidth}@{}}
                 {\small Number in
                   each cell denote number of expressions for Target that
                   can be generated from RTL patterns common with
                   Source. Numbers in parentheses denote corresponding
                   percentage for Targets.}
 \end{tabular}}
\vskip -5mm
\end{table}

Table~\ref{tab:Inter3} lists our findings. Due to its small
size, the percentage for VAX are very high if it is used as a
target, and very low if it is used as a source. Still, it is
interesting to see that only 125 RTL expressions of VAX span
771 RTL expressions of ARM. If we ignore VAX, we see that
44--69\% of the RTL expressions in the target architecture
have a similar RTL expression in the source architecture,
numbers being on the higher side for pairs of RISC
architectures. These numbers strengthen our belief that there
is lot of rework going on in retargeting GCC and it is worth
investing efforts to build an automated tool to allow reuse
of RTL patterns.

\section{Related Work}
\label{sec:related}
% \chapter{Related Work}
Several attempts have been made over last few years to
improve the process of retargeting GCC and simplifying MD
file creation.  Khedker et al., proposed a new language
specRTL~\cite{Khedker} to describe RTL expressions at an
abstract level (also called patterns) and use concrete
details to translate them to generate MD file. Sameera et al.~\cite{Sameera} proposed a systematic way to
build GCC machine descriptions. Kai-Wei Lin et al.~\cite{kai-wei} describe a systematic
methodology to port GCC to a new architecture. A detailed
description of these and more related work could be found in
Saravana's thesis~\cite{saravana2012retarget}.

\section{Conclusions}
\label{sec:concl}
% \chapter{Conclusion and Future Work}

Retargeting GCC compiler to a new architecture is a challenging
job. It can be made simpler by automating the
generation of parts of MD files, by reusing
information from existing MD files. In this
paper we showed, through empirical
studies, that popular architectures show a lot of
similarities in their MD files. This supports the observation
of Davidson and Fraser~\cite{Davidson} that it is possible to
generate machine-dependent code from a low level
machine-independent form that is common across multiple
architectures. 

As part of our work, we implemented {\em mdcompare} to study and
compare MD files. We chose to concentrate on RTL expressions
in MD files, which are used to describe machine-specific
instructions in machine-independent form in GCC. We devised a
method to extract RTL patterns from the RTL expressions to
ease the understanding and comparison. With this tool, we
were able to measure similarities between machine description
files based on the common RTL patterns. The results obtained
are promising and can be used to justify efforts to build an
automated tool to generate parts of MD file for a new
architecture using a similar architecture.

One of the immediate goal is to create a Graphical
User Interface (GUI) that can help the user to visualize RTL
patterns and experiment with it by supplying machine-specific
parameters. This will help in understanding MD files of
existing architecture. The long term goal of this work is to
build a tool which can automatically generate parts of MD
files from existing MD files with user assistance. While it
seems impossible to generate a complete MD file
automatically, we believe that a partially generated MD file
will help the user by reducing the efforts to populate the MD
file and by reducing the chances of errors which occur during
the process.

\bibliographystyle{abbrv}
\bibliography{gcc_retarget}

\end{document}